\documentclass[prl,aps,twocolumn,superscriptaddress,floats,showpacs,final, fleqn,hidelinks]{revtex4-2}
\usepackage{dcolumn}
\usepackage{bm}
\usepackage{color}
\usepackage{mathtools}
\usepackage{amsmath,amsfonts,amssymb,bm}
\usepackage{graphicx,tikz}
\usepackage[pdftex, bookmarks=true, linktoc=section, pdfstartview=FitH]{hyperref}

\newcommand{\be}{\begin{equation}}
\newcommand{\ee}{\end{equation}}
\newcommand{\ba}{\begin{eqnarray}}
\newcommand{\ea}{\end{eqnarray}}

\newcommand{\discretep}[2]{\ensuremath{P_{\{#1\}}(#2)}}

\begin{document}

\title{Optimal Local Simulations of a Quantum Singlet}

\author{David Llamas}
\affiliation{Department of Physics, University of Massachusetts Boston, Boston, Massachusetts 02125, USA}

\author{Dmitry Chistikov}
\affiliation{Centre for Discrete Mathematics and its Applications (DIMAP) \& Department of Computer Science, University of Warwick,
Coventry, CV4 7AL, U.K.}

\author{Adrian Kent}
\email{apak@cam.ac.uk}
\affiliation{Centre for Quantum Information and Foundations, Department of Applied Mathematics and Theoretical Physics, University of Cambridge, Wilberforce Road, Cambridge, CB3 0WA, U.K.}
\affiliation{Perimeter Institute for Theoretical Physics, 31 Caroline Street North, Waterloo, Ontario N2L 2Y5, Canada.}

\author{Mike Paterson}
\affiliation{Centre for Discrete Mathematics and its Applications (DIMAP) \& Department of Computer Science, University of Warwick,
Coventry, CV4 7AL, U.K.}

\author{Olga Goulko}
\email{olga.goulko@umb.edu}
\affiliation{Department of Physics, University of Massachusetts Boston, Boston, Massachusetts 02125, USA}

\date{\today}

\begin{abstract}
Bell's seminal work showed that no local hidden variable (LHV) model can fully reproduce the quantum correlations of a two-qubit singlet state.  His argument and later developments by Clauser \textit{et~al.}\ effectively 
rely on gaps between the anti-correlations achievable by classical models and quantum theory for projective measurements along randomly chosen axes separated by a fixed angle.  However, the size of these gaps has to date remained unknown.   Here we numerically determine the LHV models maximizing anti-correlations for random axes separated by any fixed angle, by mapping the problem onto ground-state configurations of fixed-range spin models.  
We identify angles where this gap is largest and thus best suited for Bell tests. These findings enrich the understanding of Bell non-locality as a physical resource in quantum information theory and quantum cryptography.
\end{abstract}

\maketitle

\paragraph{Introduction. }
%relevance for quantum information. 
In his celebrated work~\cite{Bell,Bell71,B81} on the incompatibility of quantum theory and local hidden variable (LHV) models, Bell introduced a model defining the outcomes of 
projective measurements on the Bloch sphere by oppositely colored hemispheres, with a randomly chosen symmetry axis, whose colors correspond to binary measurement outcomes.  
Modeling a two-qubit system with two opposite colorings reproduces the perfect anti-correlations and 
the rotational invariance of the singlet.    
Famously, however, Bell showed that neither this nor any other local model reproduces general quantum correlations for independently chosen measurements on the two-qubit singlet state, by analyzing measurements for 
axes separated by small angles. Clauser et al.\ \cite{CHSH69} gave an experimentally accessible measure
of Bell non-locality via the Clauser-Horne-Shimony-Holt (CHSH) inequalities, which use pairs of measurement axes separated by $\pi/4$.  
These were generalized by Braunstein-Caves \cite{BC90} to pairs separated by $\pi/2N$ for integer $N>2$. 
Experiments (e.g.\ \cite{AGR82,SBHGZ08,TBZG98,YD92.1,SMCBWSGGHACDHLVLTMZSAAPJMKBMKN15,GVMWHHPSKLAAPMBGLSNSUWZ15,HBDRKBRVSAAPMMTEWTH15}) have confirmed quantum theory and 
refuted LHV theories, with progressively fewer assumptions.  

Modern treatments consider Bell non-locality as a resource to be quantified in ways that capture its value for quantum information theoretic tasks. Much remains to be explored here.   
For example, it was only recently shown~\cite{chistikov2020globehopping} 
that the Bell hemisphere model does not optimally approximate the singlet
correlations for generic measurement axis choices: i.e., there are generally
LHV models that give anti-correlations closer to the quantum
singlet anti-correlations for randomly chosen pairs of axes separated by a given angle $\theta$.
Moreover, even if we restrict to planar polarization measurements that are parametrized by a great circle on the Bloch sphere, the optimal
approximation to singlet correlations is not always attained by models in which the two qubits have opposite colorings on the circle~\cite{chistikov2020globehopping}.
Identifying the optimal LHV model for two qubits requires us to consider models that assign
independent, and not necessarily simply related, probability
distributions on the Bloch sphere 
for each qubit's hidden variables. 

Here we address the problem of finding the optimal local models for this general task as well as the interesting restricted
case in which the colorings are opposite.   
By transforming these into ground state problems for unusual long-range spin
systems~\cite{goulko2017grasshopper}, we present numerical characterizations of the optimal models. 
Our results give a rich new quantitative characterization of the Bell
non-locality of the singlet as a fundamental physical resource.   One class of applications is in efficient testing of singlets~\cite{cowperthwaite2023comparingsinglettestingschemes} to guarantee the integrity of quantum
cryptography and other tasks in scenarios where adversaries may substitute physical systems with pre-determined measurement outcomes for qubits.  

\paragraph{Background. }
Ref.~\cite{kent2014bloch} introduced and analyzed Bell inequalities for spin measurements along two random axes that are separated by a fixed angle $\theta$. It was shown that classical correlations satisfy bounds violated by quantum theory for $\theta$ in the range $0 \leq \theta\leq \pi/3$, and that the Bell hemisphere model is not optimal for large values of $\theta$. However, the bounds were not shown to be optimal for general $\theta$ and optimal classical models were not identified. It was noted that tighter bounds and classical models closer to optimal would be obtained by solving the following intriguing geometric combinatorics problem:

A Bloch sphere is half-covered by a lawn, such that exactly one of every pair of antipodal points belongs to the lawn. A grasshopper lands at a random point on the lawn, and then jumps in a random direction through spherical angle $\theta$. What lawn shape maximizes the probability that the grasshopper remains on the lawn after jumping, and what is this maximum probability as a function of $\theta$?

It was also noted that optimal bounds and models might require solving a version of this problem with two independent overlapping lawns, in which the grasshopper jumps from one lawn and the aim is to maximize the probability that it lands outside the other.   

%prior results on the plane
Although simple to state, even the one-lawn version is a difficult mathematical problem to solve. A natural simplification is to first consider this problem on the plane, which was discussed in~\cite{goulko2017grasshopper}. It was shown analytically~\cite{goulko2017grasshopper,llamas2023grasshopper} that, despite the rotationally symmetric setup, there is no value of the jump for which a disk-shaped lawn is optimal. Optimal lawn shapes were obtained through numerical means for a range of jump lengths, via a mapping onto a statistical Ising-type model with fixed-range interactions and conserved total spin.

%summary of prior analytical results on the sphere
Ref.~\cite{chistikov2020globehopping} presented analytical results for the grasshopper problem on the circle and the sphere. This setup can be experimentally realized with projective polarization measurements on photons. For the circle, it was shown that in almost all cases lawns can be constructed such that the supremum of the probability for the grasshopper to remain on the lawn is one. In other words, LHV models can produce perfect anti-correlations. The only exception are antipodal lawns with a jump angle of the form $\theta=\pi\cdot p/q$, where $p$ and $q$ are co-prime integers, $p$ is odd, and $q$ is even, in which case the optimal probability equals $1-1/q$. 
The case with $p$ and $q$ both odd requires two independent
lawns to produce perfect anti-correlations. 

For the sphere, similarly to the planar case, hemispherical lawns are almost never optimal~\cite{chistikov2020globehopping} (except for jump angles of the form $\theta=\pi/q$ for integer $q>1$). This means that more sophisticated LHV models can be constructed for generic $\theta$, leading to novel types of Bell inequalities.  Numerical investigation of the actual shapes of these optimal non-hemispherical lawns is the focus of this work.

\paragraph{Model and Methods. }
LHV models defined by binary Bloch-sphere colorings can be characterized by the lawn function $\mu(\mathbf r)\in\{0,1\}$ for $\mathbf{r}\in\mathbb{S}^2$, where $\mu(\mathbf r)=1$ denotes points on the lawn and $\mu(\mathbf r)=0$ points not on the lawn (say, spin up and spin down, respectively). For a unit sphere, spherical distances equal spherical angles, so the normalization is $\int_{\mathbb{S}^2}d^2\mathbf{r}\mu(\mathbf r)=2\pi.$ The antipodal condition implies $\mu(\mathbf r)+\mu(-\mathbf r)=1$ for all $\mathbf r\in\mathbb{S}^2$.

In the general two-qubit setup, $L_1$ and $L_2$ denote the lawns on the two independent Bloch spheres, while $\overline{L}_1$ and $\overline{L}_2$ denote their complements, i.e., $L_k$ ($\overline{L}_k$) is the set of all points on the sphere such that $\mu_k(\mathbf r)=1$ ($\mu_k(\mathbf r)=0$). In the special case of the singlet, the colorings of the two Bloch spheres, and thus the lawns, are complementary, $L_1=\overline{L}_2$. In this scenario, the two lawns have opposite colorings and the grasshopper aims to remain on the same lawn after jumping. We call this the ``one-lawn setup", because there is no need to consider the two complementary lawns separately. The general setup of two independent lawns, where the grasshopper aims to jump from $L_1$ onto $\overline{L}_2$ without the additional restriction, is also called the ``two-lawn setup".
\begin{figure*}
    \centering
    \includegraphics[width=\textwidth]{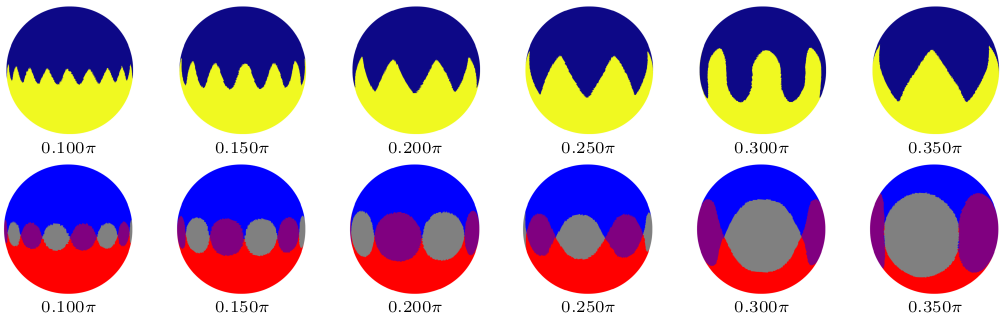}
    \caption{Examples of optimal grasshopper lawn shapes in the cogwheel regime for the one-lawn (top panel) and two-lawn (bottom panel) setups. In the two-lawn setup, the two lawns are denoted by the red and blue colors. Gray areas are not covered by either lawn and purple areas are covered by both lawns. 
    }
    \label{fig:both-cogwheels}
\end{figure*}

The grasshopper success probability for jump angle $\theta$ is then given by
\begin{equation}
    P(\theta)=
    \frac{1}{4\pi^2\sin(\theta)}\int_{L_1}d^2r_1\int_{\overline{L}_2}d^2r_2\delta(\theta_{12}-\theta),\label{eq:continuous}
\end{equation}
where $\theta_{12}$ is the spherical angle between the points $\mathbf{r}_1$ and $\mathbf{r}_2$ on the unit sphere.

Determining optimal grasshopper lawn shapes is a complex global optimization problem. To treat it numerically, we discretize it by imposing an antipodal grid with $N$ points on the unit sphere. Exactly one of a pair of opposite poles is occupied by the lawn. The results presented here were obtained for spherical $t$-design grids~\cite{Womersley2017grids, Womersley2018spheregrid} with $N = 52978$ sites, which provide a good approximation of the continuous sphere with discretization errors below $0.1\%$. A detailed discussion of different grid setups and corresponding accuracy tests are provided in the companion paper~\cite{llamas2024sphericallongpaper}. 

The discretization procedure is analogous to the planar case~\cite{goulko2017grasshopper}.
In the one-lawn setup, the success probability~\eqref{eq:continuous} maps onto
\begin{equation}
\discretep{s}{\theta} = \frac{4}{\sin(\theta) N^2 h}\sum_{i, j}  s_i s_j \phi \left( \frac{ \theta_{ij} - \theta}{h} \right), \label{eq:spinham}
\end{equation}
where $s_i=\mu(\mathbf r_i)$ is the lawn function at grid site $i$, $h=\sqrt{4\pi/N}$ is the average lattice spacing on the spherical grid, and $\phi(r)$ is a suitable discrete approximation of the Dirac $\delta$-function, $\delta(r)\rightarrow\delta_h(r)=\phi(r/h)/h$~\cite{peskin2002deltafn}. 
Normalization implies $\sum_i s_i = N/2 = 2\pi/h^2$. With this notation, the problem can be interpreted as a conserved-spin Ising model with Hamiltonian $H = -\discretep{s}{\theta}$ on a spherical grid, where $s_i\in\{0,1\}$ are the Ising spins. In contrast to conventional Ising models, this system has long-range (and fixed-range) interactions. The interaction distance corresponds to the grasshopper jump angle $\theta$, which is resolved by a large number of grid spacings, $h\ll\theta$. In the continuum limit, $h\rightarrow 0$ and $N\rightarrow\infty$, the discrete $\discretep{s}{\theta}$ approaches the continuous $P(\theta)$.

The setup with two independent lawns can be treated analogously. In this case, two discrete lawn configurations need to be accounted for, with $s_i$ belonging to the first discrete lawn and $s_j$ to the complement of the second discrete lawn, respectively.

We use an efficient global optimization protocol based on simulated annealing to find the optimal grasshopper lawn shapes (which correspond to ground states of the discrete Ising model). Details are provided in the companion paper~\cite{llamas2024sphericallongpaper}.

\paragraph{Results. }
While it has been previously known~\cite{chistikov2020globehopping} that hemispherical lawns cannot be optimal solutions to the one-lawn spherical grasshopper problem for almost any value of $\theta$, the specific shapes of the optimal solutions and the associated probabilities remained open questions. Here we present our numerical results for these quantities and interpret them in the quantum information context.
\begin{figure}
    \centering
    \includegraphics[width=\columnwidth]{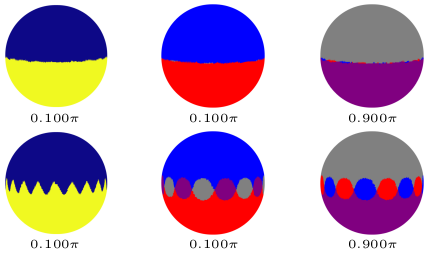}
    \caption{The jump $\theta=0.1\pi$ is of the form $\theta_{q=10}=\pi/10$ for even $q=10$. There are two classes of near-optimal solutions in this case, the hemisphere (top) and the cogwheel (bottom). The images on the left are for the one-lawn setup and the remaining images for the two-lawn setup, shown for both $\theta=\theta_{10}$ (center) and $\theta=\pi-\theta_{10}=0.9\pi$ (right).}
    \label{fig:cog-hemisphere}
\end{figure}

%cogwheel regime
For jump angles $\theta\lesssim 0.41\pi$, numerically found optimal lawns resemble cogwheels in both the one-lawn and two-lawn setups (see Fig.~\ref{fig:both-cogwheels} for examples), as in the planar problem for jump distances $d \lesssim 0.57$. 
Due to the antipodal condition, the number of cogs per lawn must be odd.
In the planar problem, the optimal cog number was found to be well-described by the number of edges of a polygon inscribed in the unit-area disk, such that the polygon edge length is close to the grasshopper jump length~\cite{goulko2017grasshopper, llamas2023grasshopper}. Higher modes with more cogs were found to give local optima~\cite{llamas2023grasshopper}. The spherical grasshopper problem exhibits similar behavior, with the optimal number of cogs in the one-lawn setup found to be the odd integer nearest to $2\pi/\theta$, which implies that the distance between the cogs approximates the jump length. Higher modes, i.e.\ odd integers nearest to $4\pi/\theta$, $6\pi/\theta$, and so on, represent local optima of the problem. 

A similar picture emerges in the two-lawn setup. Here the grasshopper endeavors to jump from one lawn $L_1$ onto the complement of the other, $\overline{L}_2$, so the cogwheel-shaped optimal lawns, while having the same shape and cog number, are typically offset by half the cog spacing, and the optimal cog number is the odd integer nearest to $\pi/\theta$. Hence for the same jump angle, the first mode of the two-lawn setup has roughly half as many cogs as the first one-lawn mode (the antipodal restriction that the cog number for each lawn must be odd remains in effect). Higher modes are also observed in the two-lawn setup, with cog numbers approximated by $2\pi/\theta$, $3\pi/\theta,\ldots$ For even-numbered modes, the lawns are not offset but exactly complementary, so that the configurations are identical to the ones in the one-lawn setup. Higher modes in both setups are discussed in more detail in the companion paper~\cite{llamas2024sphericallongpaper}.
\begin{figure*}
    \centering
    \includegraphics[width=\textwidth]{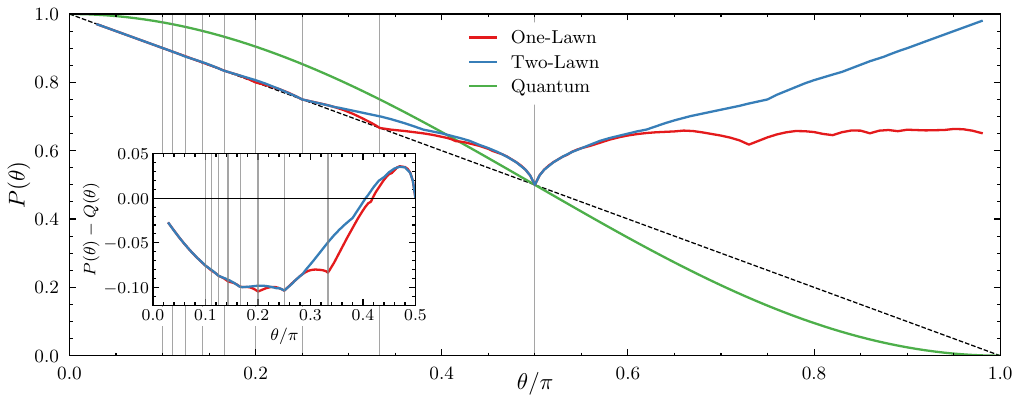}
    \caption{Probability of optimal lawn shapes as a function of $\theta$ in the one-lawn setup (red solid line) and the two-lawn setup (blue solid line). The quantum probability of anti-correlations for a quantum singlet state, $Q(\theta)$, is shown for comparison (green solid line). The diagonal dashed black line corresponds to the probability for a hemispherical lawn, $1-\theta/\pi$. Vertical gray lines denote jump angles of the form $\theta=\theta_q$ for integer $q$. The inset shows the difference between the classical and quantum correlations in the two setups.}
    \label{fig:probabilities}
\end{figure*}

It was shown in~\cite{chistikov2020globehopping} that the hemisphere is an optimum of the one-lawn version of the grasshopper problem if and only if the jump angle is $\theta=\theta_q\equiv\pi/q$, for integer $q>1$. Indeed numerical results confirm that for angles $\theta_q$ the hemisphere is at least near-optimal. However, numerical searches for an optimal configuration also produce at least one cogwheel configuration, significantly different from the hemisphere, with essentially the same probability. In other words, the maximum appears to be near-degenerate for these jumps, see Fig.~\ref{fig:cog-hemisphere}. 

For two independent lawns, it was shown in~\cite{kent2014bloch} that the hemisphere is optimal for angles of the form $\theta_q$ when $q$ is even, and
in~\cite{chistikov2020globehopping} that it is not optimal for any other angles.  Intuitively, we can offset the two lawns, such that the smallest optimal number of cogs is approximated by $\pi/\theta$, rather than by $2\pi/\theta$ as in the one-lawn setup. For $\theta=\theta_q$, this translates to approximately $q$ and $2q$ cogs for the two-lawn and one-lawn setups, respectively. When this number is even, the actual number of cogs must be rounded to the nearest odd integer, which is the ``worst case scenario" in terms of the extent of rounding required. In these cases the cogwheel solutions fail to outperform the hemispherical solutions. In the one-lawn setup this happens for every integer $q$, as $2q$ is always even. In the two-lawn setup on the other hand, for odd values of $q$ no rounding is required as $\pi/\theta_q$ is an odd integer. Thus the cogwheel solution has a higher success probability than the hemispherical one in these cases.

In the two-lawn setup, the cogwheel regime recurs for angles $\theta\gtrsim 0.59\pi = \pi-0.41\pi$ because for jumps $\theta$ and $\pi-\theta$ the optimal shapes are the same, except that one of the lawns is inverted. Explicitly, if $L_1$ and $L_2$ are optimal for an angle $\theta$ below $\pi/2$, then $L_1$ and $\overline{L}_2$ (or equivalently $\overline{L}_1$ and $L_2$) are optimal for $\pi-\theta$. This follows from the symmetry of the probability integral \eqref{eq:continuous} in the general two-lawn case. An example can be seen in Fig.~\ref{fig:cog-hemisphere} for $\theta=0.1\pi$ and $\theta=0.9\pi$.

%briefly discuss other regimes
In the one-lawn setup, because of the antipodal condition, for $\theta=\pi$ all lawns have success probability zero, regardless of their shape. For the same reason, for $\theta=\pi/2$ all lawns have success probability $1/2$ (the set of possible landing points is a great circle exactly half of which is covered by the lawn). In contrast, in the two-lawn setup, due to the $\theta\leftrightarrow\pi-\theta$ symmetry the optimal shapes for angles near $\pi$ are the same as for angles near $0$ and the grasshopper success probability approaches 1 as $\theta\rightarrow \pi$. Near $\pi/2$ the probability approaches $1/2$, just as in the one-lawn setup.
Numerically found optimal grasshopper shapes in the vicinity of $\theta=\pi/2$ are labyrinth-like for both setups. 
For $\theta\gtrsim0.57\pi$ in the one-lawn setup the configurations become striped, with cog-like modulations along the stripe boundaries that become less pronounced as $\theta$ increases. The stripe width decreases with increasing $\theta$ and as $\theta$ approaches $\pi$, optimized configurations develop defects in otherwise regular stripe patterns. 
For images and a detailed discussion of the labyrinth and stripe regimes we refer to the companion paper~\cite{llamas2024sphericallongpaper}.

\paragraph{Discussion. }
We have numerically established the nature of the optimal lawn shapes across all $0\leq\theta\leq\pi$ for both the complementary (one-lawn) and the independent (two-lawn) setups. We reproduced the analytical results that hemispherical lawns are not optimal, unless $\theta=\pi/q$ for integer $q$ in the one-lawn setup, and even integer $q$ in the two-lawn setup, respectively. In these special cases, hemispherical lawns are indeed optimal, but there are also significantly different cogwheel shapes with very similar probability. For generic $\theta\lesssim0.41\pi$  optimal configurations are cogwheels for both setups; the symmetry of the independent lawn setup implies the same is true there for $\theta\gtrsim0.59\pi$. We also identified optimal shapes for other angles.   

The grasshopper success probabilities quantify the correlations of the corresponding LHVs. The probabilities $P(\theta)$ for the one- and two-lawn setups are shown in Fig.~\ref{fig:probabilities} and are compared with the quantum singlet probability $Q(\theta)=\cos^2(\theta/2)$. 
Quantum anti-correlations are stronger than those given by optimal lawns for $\theta\lesssim 0.41\pi$ (approximately the range of the cogwheel regime, though the precise ranges for the one- and two-lawn setups differ slightly).
For jumps below $\pi/2$, the largest difference between the quantum singlet and the optimal LHV for independent lawns occurs at $\pi/4$, the angle separating axes in a CHSH non-locality test~\cite{CHSH69}, taking the value 
$\cos^2 (\pi/8) - 3/4 \approx 0.10355$.  
For complementary lawns the largest gap is $\cos^2 ( \pi /10) - 4/5 \approx 0.10451$ at $\pi/5$. The numerically obtained values for these gaps agree with theory within 0.2\%, confirming numerically that the complementary lawn gap is larger at $\pi/5$ than at $\pi / 4$.  To understand these results, note that the gap between the Bell hemispherical model and quantum correlations is maximized at $\theta= \pi/n$, where $n = \pi/ {\arcsin (2/ \pi )} \approx 4.5523$.   However, for $4<n<5$ the optimal LHV one- and two-lawn models are given by cogwheel rather than hemispherical lawns, and this ``cogwheel advantage" means the gap is not maximized in this region.  For independent lawns the optimal lawns are hemispherical and complementary at $\pi/4$ but not $\pi/5$~\cite{kent2014bloch}, while for complementary lawns the optimal lawn is hemispherical at $\pi/5$~\cite{chistikov2020globehopping}.   These two cases give the optimal gaps among angles where the optimal lawn is hemispherical. 
      
Optimally efficient discrimination between singlets and LHVs in the class of tests considered here can thus be achieved by measurements of spin about random axes separated by $\pi/4$.   Because CHSH tests using axes defined in any great circle have the same efficiency, this yields an important corollary: {\it any} CHSH test achieves the optimal efficiency attainable by the much larger class of random axis measurement tests considered here. 
   
Optimally efficient discrimination tests between singlets and anti-correlated LHVs are given by measuring spin about random axes separated by $\pi/5$. As far as we are aware, this non-locality test has not previously been considered in the literature.  Our result implies it is optimal among the class we consider in scenarios where adversaries are constrained to produce anti-correlated LHVs, either by limited physical resources, or perhaps because there is a very large penalty for failing an anti-correlation test (which is the only deterministic test involving local measurements on a singlet).  
It also motivates exploring mixtures of anti-correlation tests and $\pi/5$ separation random axis measurements, which are candidates for a more efficient test of the singlet against general LHVs. 

It would be very interesting to investigate optimality for these and other mixed tests involving two or more angles.  For example, a suitable mixture of $\pi/4$ and $\arcsin (2/\pi )$ tests also might be more efficient than either, given that the optimal lawn for $\pi/4$ tests is hemispherical and the gap between quantum singlet correlations and those of the hemispherical lawn model is larger at $\arcsin (2/\pi)$.  Extending our methodology to general mixed tests is a task for future work.

\paragraph{Acknowledgements. }
We thank Carlo Piovesan for sharing unpublished numerical results. 
This work is supported by the NSF under Grant No. PHY-2112738 and Grant No. OSI-2328774 (O.G. and D.L.). O.G. also acknowledges support under NSF Grant No. PHY-2441282. The work of D.L. was supported in part by College of Science and Mathematics Dean's Doctoral Research Fellowship through fellowship support from Oracle, project ID R0000000025727.
A.K. acknowledges financial support from project OPP640, funded by the Science and Technology Facilities Council's International Science Partnerships Fund.
A.K. was supported in part by Perimeter Institute for Theoretical Physics. Research at Perimeter Institute is supported by the Government of Canada through the Department of Innovation, Science and Economic Development and by the Province of Ontario through the Ministry of Research, Innovation and Science. 
D.C. and M.P. acknowledge support from DIMAP and the Department of Computer Science at the University of Warwick.

\bibliography{grasshopper,postdocbiblio}

\end{document}